\documentclass[conference]{IEEEtran}
\IEEEoverridecommandlockouts
\usepackage[T1]{fontenc}
\usepackage{enumitem}
\usepackage{amsmath}
\usepackage[cmintegrals]{newtxmath}
\usepackage{cite}
\usepackage{amsmath,amsfonts}
\usepackage{algorithm}  
\usepackage{algpseudocode}
\usepackage{graphicx}
\usepackage{textcomp}
\usepackage{xcolor, amsmath}
\usepackage{amsmath}
\usepackage{bbm}
\usepackage{bbold}
\usepackage{bm}
\usepackage{color}
\usepackage{soul}
\usepackage{hyperref}
\usepackage{caption}
\usepackage{booktabs}
\usepackage{longtable}
\usepackage{tabularx}
\usepackage{caption}
\def\BibTeX{{\rm B\kern-.05em{\sc i\kern-.025em b}\kern-.08em
    T\kern-.1667em\lower.7ex\hbox{E}\kern-.125emX}} 
\begin{document}
\title{Hybrid SMI Realization via Matrix Completion and Riemannian Manifold Optimization on Narrowband Sub-Array Based Architectures\\
}
\author{
\IEEEauthorblockN{T. S. Cousik, R. Rangaraj, N. Tripathi, J. H. Reed}
\IEEEauthorblockA{
\textit{Wireless@VT} \\
Virginia Tech, Blacksburg, USA \\
\{tarunsc, rrohit, nishith, reedjh\}@vt.edu
}
\and
\IEEEauthorblockN{D. J. Jakubisin}
\IEEEauthorblockA{
\textit{National Security Institute} \\
Virginia Tech, Blacksburg, USA \\
djj@vt.edu
}
\and
\IEEEauthorblockN{J. Kraft}
\IEEEauthorblockA{
\textit{Analog Devices} \\
Denver, USA \\
jon.kraft@analog.com
}
}
\maketitle

\begin{abstract}Hybrid beamforming architectures reduce hardware complexity but restrict access to full-array observations, rendering direct implementation of classical covariance-based methods such as minimum variance distortionless response (MVDR) and sample matrix inversion (SMI) infeasible. This work introduces a structured covariance completion framework, termed Rock Road to Dublin (RR2D), which estimates the unobservable analytical covariance matrix (ACM) from a partially observed sample covariance matrix (SCM). RR2D exploits signal stationarity across the array and enforces physical measurement consistency using Dykstra’s alternating projection algorithm with positive semidefinite, Toeplitz, and block constraints. The reconstructed virtual ACM enables a realizable hybrid SMI (H-SMI) formulation that remains fully compatible with existing hybrid MVDR optimization frameworks. Empirical results for a 32-element hybrid array demonstrate both the expected degradation of H-SMI implemented directly under prior H-MVDR formulations and the performance gains achieved through RR2D. The proposed H-SMI consistently outperforms previous hybrid SMI and partial digital baselines, achieving performance close to the  H-MVDR reference. Overall, RR2D bridges the gap between theoretical H MVDR formulations and practical hybrid hardware by enabling structured covariance reconstruction from incomplete observations.
\end{abstract}
\section{Introduction and Motivation}
Radar and wireless communication systems employ adaptive beamforming to dynamically adjust array weights based on received data, optimizing objectives such as minimizing noise variance, maximizing SINR, or minimizing mean square error (MSE) relative to a reference signal.\cite{VanTrees-OAP}. 
To achieve these goals, adaptive arrays exploit instantaneous or statistical characteristics of the environment to form data-driven spatial filters. Among existing adaptive beamforming structures, the minimum variance distortionless response (MVDR) beamformer minimizes output noise power while preserving the signal of interest (SOI) in a distortionless manner\cite{VanTrees-OAP}. Assuming the noise is sampled from a Guassian random process, the output of MVDR forms the maximum likelihood estimate of the SOI. MVDR relies on knowledge of the analytical covariance matrix (ACM), which in practice is not feasible and must be replaced by its sample estimate, the sample covariance matrix (SCM), which produces the sample matrix inversion (SMI) formulation \cite{VanTrees-OAP}. Seminal work in \cite{RMB-Bound} showed that when the number of snapshots is approximately twice the number of array elements, the SMI beamformer achieves SINR performance that is within  3~dB of the ideal MVDR solution. SMI was also shown to achieve the SINR performance of MVDR in the asymptote. 
SMI and MVDR estimate the SCM across all array elements, which is then subsequently used to adapt the weights. In practice, this typically necessitates connecting each array element to an independent 
analog-to-digital conversion (ADC) chain to realize MVDR or SMI.  Digital beamformers (DBFs) offer the highest degrees of freedom by 
digitizing every antenna element, but their power, cost, and 
synchronization overhead scale poorly with array size. As array apertures and 
bandwidths increase, the cost and power demands of DBFs 
become a primary constraint, motivating alternative receiver architectures that trade degrees of freedom for reduced system complexity and cost.

Hybrid beamformers 
(HBFs) reduce this cost by combining processing across analog and digital channels; \textit{typically} 
the number of digital channels is smaller than the number of antenna 
elements. In Sub-Array-Based Hybrid beamformers (S-HBFs), each digital 
channel is connected to a subset of antenna elements, called a sub-array. 
When each sub-array of a S-HBF contains only one antenna element, it is called a partial digital beamformer (P-DBF), and it serves as a convenient 
performance baseline.

In the context of spatial multiplexing for multi-user MIMO system\cite{H-MIMO}, showed that HBFs can match DBF performance if the number of RF chains was double the number of digital streams. Recent work in \cite{fc-mvdr} -\cite{h-mvdr} introduced Hybrid MVDR (H-MVDR) formulations for 
fully connected and sub-array based architectures respectively. The work in 
\cite{fc-mvdr} employed sparse recovery to approximate the 
H-MVDR solution for fully connected HBFs (FC-HBFs), while  
\cite{h-mvdr} compared both sparse recovery and Riemannian manifold optimization approaches in S-HBFs to address 
the unit-modulus constraint on the analog weights. 

\begin{figure*}[!t]
    \centering
    \includegraphics[
        width=\textwidth,
        height=2.2in,
        keepaspectratio=True,
        clip,
        trim=10pt 250pt 50pt 25pt
    ]{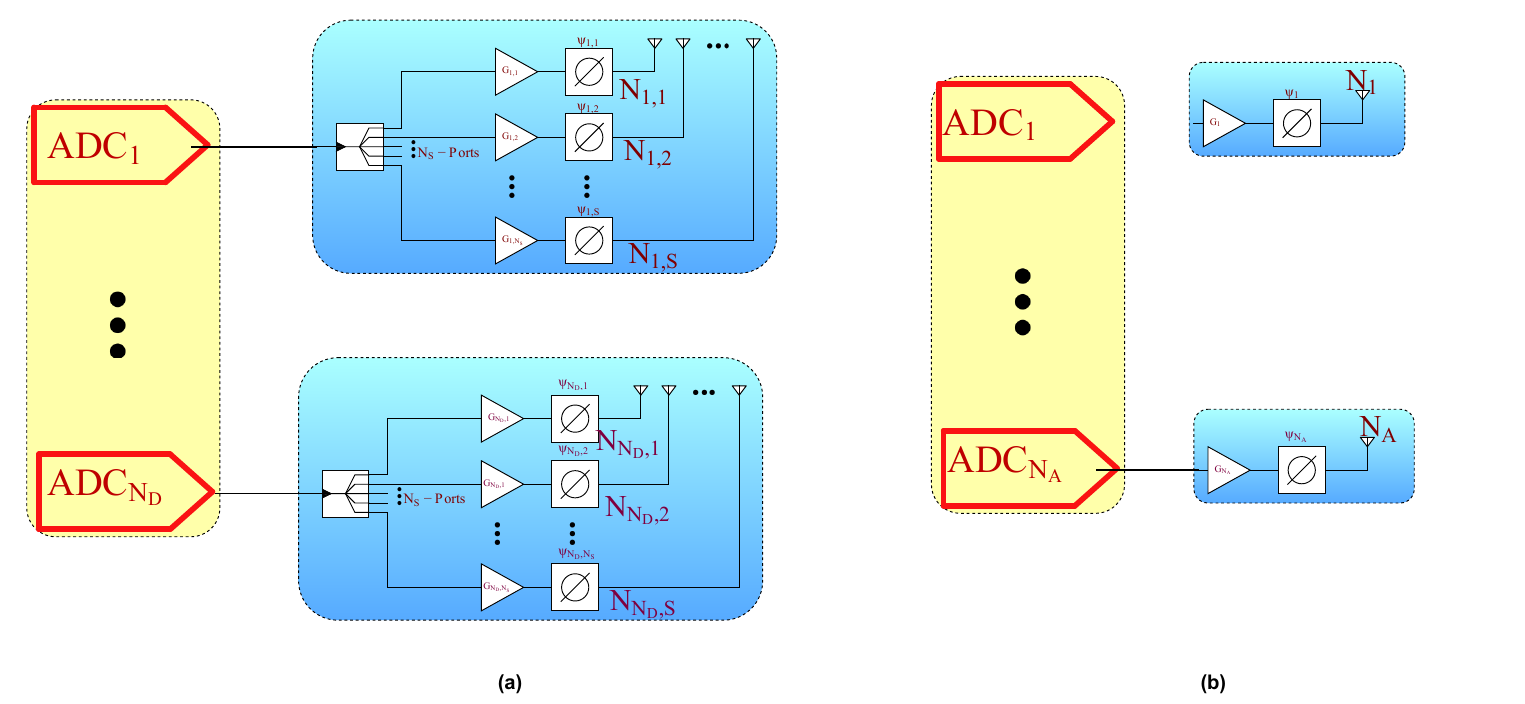}
    \caption{Figure 1(a) illustrates the S HBF architecture considered in this work. Digital chains that feed the baseband processor are shown in yellow boxes, and the blue regions indicate the analog and RF channels in each subarray. Figure 1(b) shows the DBF architecture used as a reference. Optional RF phase and gain control blocks are included for generality although they are not used in the presented results.}
    \label{fig:SHBF_vs_DBF}
\end{figure*}

Prior work in \cite{fc-mvdr} demonstrated that FC-HBFs can match DBFs, and that near 
\cite{h-mvdr} DBF performance can be obtained using S-HBFs at a fraction of the design cost and complexity. 
Figures \ref{fig:SHBF_vs_DBF} present the exemplary S-HBF and DBF array architectures that are used as reference models throughout this manuscript.

These studies \cite{fc-mvdr}-\cite{h-mvdr} represent an important step toward practical Hybrid SMI (H-SMI) 
implementations. Nevertheless, the current formulations in  \cite{fc-mvdr}-\cite{h-mvdr} benefit from 
further development and validation in several areas to enable full reduction to practice, these include:
\vspace{-3 pt}
\begin{enumerate}[leftmargin=*]
\item \textbf{Dependence on unobservable covariances:} Both studies \cite{fc-mvdr}-\cite{h-mvdr} require access to the complete ACM of the array in order to compute the MVDR weights of a auxiliary DBF. The calculated weights of the auxiliary DBF are subsequently used in determining the analog and digital weights for the HBF. In practical HBFs, only the 
SCM is available and moreover the SCM's dimension is limited to 
the number of digital channels, which is lower than the number of array elemenets. Consequently, the performance reported in \cite{fc-mvdr}-\cite{h-mvdr} are limited to reflecting an upper bound that, in their current form, cannot be achieved directly on hardware.

\item \textbf{Lack of covariance reconstruction framework:} Neither formulation proposes a mechanism to estimate or recover the unobserved covariance information from hybrid observations. The bridge between 
theoretical and implementable hybrid architectures, therefore, remains incomplete. A reconstruction approach is needed to infer an equivalent analog-domain covariance that maintains the spatial structure of the true array's statistics.

\item \textbf{Covariance estimation inconsistency:}
In prior works \cite{fc-mvdr}-\cite{h-mvdr}, the covariance matrix was computed from the 
total received signal, which includes the SOI component. When the desired 
signal contributes to the covariance, it is represented 
in the precision matrix, leading to undesired nulling of the SOI in the MVDR response, particularly at moderate and high 
SNRs. In practice, the covariance used for MVDR or SMI beamforming should be estimated using interference-plus-noise snapshots only, excluding the SOI through pre-filtering. 

\item \textbf{Restricted evaluation conditions:} 
The analyses in \cite{h-mvdr}–\cite{fc-mvdr} were limited to a single array geometry and fixed source configuration, motivating further study under varying geometries and diverse signal and interference conditions.

\item \textbf{Absence of statistical normalization:} Reported SINR results were presented for individual scenarios without normalization or consistent statistical baselines. In this work, weight normalization and an input SINR baseline are introduced to enable fair comparison across architectures and conditions.
    
\item \textbf{Idealized interpretation of hybrid performance:} Prior studies conclude that HBFs can approach the performance 
of DBFs at lower cost. When realistic SCM estimation and reconstruction errors are considered, achievable SINR levels are lower and these performance losses must be quantified prior to practical implementation.

\item \textbf{Simplistic treatment of nonconvexity:}
The prior work identifies non-convexity as arising primarily from the 
unit-modulus constraint on the analog phase shifters, but does not analyze 
other contributing factors such as the coupling between analog and digital 
weights or the covariance inverse. The optimization landscape therefore 
remains uncharacterized, and its impact on convergence and solution robustness is not discussed. Empirical analysis of the optimization Hessian reveals a 
    mix of positive and negative eigenvalues, confirming the nonconvex 
    nature of the problem and justifying the use of manifold-based 
    optimization methods.
    
\end{enumerate}
Thus, while prior results are promising and establish a foundation for H-MVDR, realizing these methods on practical S-HBF architectures requires (i) reconstructing higher-dimensional covariances from limited observations and (ii) conducting broader statistical evaluations.

The limitations 
are discussed briefly here to motivate the proposed approach, and each is 
revisited in greater depth in subsequent sections, where their mathematical 
structure and practical implications are examined in detail.

This work addresses these needs by reconstructing a virtual 
analog-domain covariance from the observed hybrid covariance using a 
matrix-completion framework that exploits the Toeplitz structure inherent 
to linear narrowband (less than \(10\%\) fractional bandwidth) arrays. The recovered estimate enables a practically  realizable H-SMI formulation that bridges the gap between theoretical 
H-MVDR methods and practical implementations. The proposed approach is validated through extensive simulations comparing MVDR/SMI performance on DBFs, P-DBFs, and S-HBF configurations across randomized angular and 
SNR/INR conditions. 

The remainder of this paper is organized as follows. Section~II presents 
the S-HBF model, introduces H-MVDR weight optimization using manifold optimization techniques and associated limitations. Section~III 
details the implementation H-SMI through a novel covariance completion framework. Section~IV provides numerical results and comparative analysis. Section ~V provides an overview of the current limitations and outlines future scope of work. Section~VI provides a summary of the work presented in this paper.
\textbf{This is a revised preprint of a paper accepted to 2026 IEEE AESS Radar Conference. The final version will appear in IEEE Xplore}

\section{Hybrid Array Model and Signal Formulation}
\subsection{Sub-Array-Based Hybrid Structure}
We assume a S-HBF comprised of a uniform linear array (ULA) where each digital channel is 
connected to a sub-array of \( N_S \) antennas, producing a block-diagonal 
analog weight matrix \( \mathbf{W_A} \in \mathbb{} \) defined as follows:
\begin{equation}
\mathbf{W_A} =
\begin{bmatrix}
\tilde{\mathbf{w}}_1^{T} & \mathbf{0}      & \cdots & \mathbf{0} \\
\mathbf{0}               & \tilde{\mathbf{w}}_2^{T} & \cdots & \mathbf{0} \\
\vdots                   & \vdots          & \ddots & \vdots \\
\mathbf{0}               & \mathbf{0}      & \cdots & \tilde{\mathbf{w}}_{N_D}^{T}
\end{bmatrix},
\tilde{\mathbf{w}}_a^{T} \in \mathbb{C}^{1 \times N_S},a = 1, 2,\ldots, N_D
\end{equation}

Here, \( N_A = N_S N_D \) denotes the total number of antenna elements, 
and each sub-array of \( N_S \) antennas is connected to one of the 
\( N_D \) digital channels. The vector \( \tilde{\mathbf{w}}_a \) 
contains the analog amplification and phase shifts applied within the \( a\)-th sub-array, 
while the remaining entries are zeros, reflecting that sub-arrays are 
electrically disjoint. We will assume that the RF/analog chains apply unit-magnitude weights (i.e only the phase shifts are adapted), satisfying 
\( |\tilde{\mathbf{w}}_a| = \mathbf{1}_{N_S} \), where 
\( \mathbf{1}_{N_S} \) is an \( N_S \)-dimensional vector of ones, to stay consistent with prior work \cite{fc-mvdr} - \cite{h-mvdr}. 
\vspace{-3 pt}
The analog combining matrix \( \mathbf{W_A} \) operates in the RF domain, 
prior to analog-to-digital conversion, while the digital beamforming 
weights \( \mathbf{w_D} \in \mathbb{C}^{N_D \times 1} \) are applied 
in baseband after the ADC stage. The overall composite response for an 
incoming signal snapshot \( \mathbf{x} \in \mathbb{C}^{N_A \times 1} \) 
is given by
\begin{equation}
y = 
\mathbf{w_D}^{H} 
\mathbf{W_A} 
\Big(
\mathbf{a}(\theta_s) s
+ \sum_{i=1}^{N_I} \mathbf{a}(\theta_i) u_i
+ \mathbf{n}
\Big),
\label{eq:hybrid_response}
\end{equation}
where \( s \) is the signal-of-interest (SOI) arriving from direction 
\( \theta_s \), \( u_i \) is the \( i\)-th interference component arriving 
from direction \( \theta_i \), and \( \mathbf{n} \) denotes additive 
white Gaussian noise. The steering vector \( \mathbf{a}(\theta) \in 
\mathbb{C}^{N_A \times 1} \) characterizes the phase progression across 
the array, while \( N_I \) represents the number of interferers. The 
matrix \( \mathbf{W_A} \) performs analog spatial combining before 
digitization, and the digital weights \( \mathbf{w_D} \) provide 
fine-grained adaptive control in the digital domain.

\subsection{Hybrid MVDR Formulation}
\label{sec:hmvdr_formulation}

Following the formulation introduced in 
\cite{h-mvdr}\cite{fc-mvdr}, the H-MVDR beamformer seeks 
to minimize the output noise power while maintaining a distortionless 
response in the SOI's direction. The joint optimization over the analog and 
digital weights is expressed as

\begin{equation}
\min_{\mathbf{W_A}, \mathbf{w_D}}
\ \mathbf{w_D}^{H}\mathbf{W_A}\mathbf{R_A}\mathbf{W_A}^{H}\mathbf{w_D},
\quad
\text{s.t. } 
\mathbf{w_D}^{H}\mathbf{W_A}\mathbf{a}(\theta_s) = 1,
\label{eq:hybrid_mvdr}
\end{equation}

where \( \mathbf{R_A} \in \mathbb{C}^{N_A \times N_A} \) is the  ACM of SOI, noise and interference emsemble seen at the receiver, the  
\( \mathbf{a}(\theta_s) \) denotes the SOI steering vector. The closed-form optimal digital weights 
for a given analog configuration are given by

\begin{equation}
\mathbf{w_D} =
\frac{(\mathbf{W_A}\mathbf{R_A}\mathbf{W_A}^{H})^{-1}
\mathbf{W_A}\mathbf{a}(\theta_s)}
{\mathbf{a}^{H}(\theta_s)
\mathbf{W_A}^{H}(\mathbf{W_A}\mathbf{R_A}\mathbf{W_A}^{H})^{-1}
\mathbf{W_A}\mathbf{a}(\theta_s)} .
\label{eq:digital_weight}
\end{equation}

The overall optimization problem to calculate optimal analog and digital weights for H-MVDR can be shown to be:
\begin{equation}
    \max_{\mathbf{W_A}} 
    \, \mathbf{a}(\theta_0)^{H} 
    \mathbf{W_A}^{H}
    \left( \mathbf{W_A} \mathbf{R}_T \mathbf{W_A}^{H} \right)^{-1} 
    \mathbf{W_A} \mathbf{a}(\theta_0)
    \label{eq:mvdr_joint_opt}
\end{equation}

The optimization problem required to solve \eqref{eq:mvdr_joint_opt} was deemed nonconvex due to the constant-modulus 
constraint on the elements of \( \mathbf{W_A} \) \cite{h-mvdr}. Equation
\eqref{eq:hybrid_mvdr} was subsequently reformulated in \cite{h-mvdr} to enable tractable optimization, as 
outlined in next subsection.
\subsection{Riemannian Manifold Optimization on a Reformulated H-MVDR Problem}
\label{sec:hmvdr_manifold}

To make the optimization in \eqref{eq:mvdr_joint_opt} tractable, 
\cite{h-mvdr} introduced an auxiliary DBF whose weights 
\( \mathbf{w_0} \in \mathbb{C}^{N_A \times 1}\) are computed using \( \mathbf{R_A} \) (\(\mathbf{R_A} \) captures the signal, interference, and noise characteristics) and is given by
\begin{equation}
\mathbf{w_0} =
\frac{\mathbf{R_A}^{-1}\mathbf{a}(\theta_s)}
{\mathbf{a}^H(\theta_s)\mathbf{R_A}^{-1}\mathbf{a}(\theta_s)}.
\label{eq:w0_virtual_mvdr}
\end{equation} The 
HBF weight optimization problem was then recast as minimizing the Euclidean distance between the ideal DBF and HBF weights:

\begin{equation}
\min_{\mathbf{W_A}, \mathbf{w_D}} 
\| \mathbf{w_0} - \mathbf{W_A}^H \mathbf{w_D} \|_2^2 .
\label{eq:mse_formulation}
\end{equation}

For a given set of analog weights \( \mathbf{W_A} \), the corresponding 
digital beamformer was shown to be
\begin{equation}
\mathbf{w_D} = \frac{1}{N_S}\mathbf{W_A}\mathbf{w_0}.
\label{eq:wd_given_WA}
\end{equation}

The expression in \eqref{eq:wd_given_WA} can be used to obtain an expression for the objective function of \eqref{eq:mse_formulation} that depends only on \( \mathbf{W_A} \):
\begin{equation}
\max_{\mathbf{W_A}}\ 
\mathbf{w_0}^H \mathbf{W_A}^H \mathbf{W_A}\mathbf{w_0}.
\label{eq:compact_WA}
\end{equation}

By reorganizing terms, this optimization can be equivalently expressed as
\begin{equation}
\max_{\mathbf{w_a}}\ 
\mathbf{w_a}^H \mathbf{W_0}^H \mathbf{W_0}\mathbf{w_a}
\quad \text{s.t.}\quad |[\mathbf{w_a}]_k| = 1,\ \forall k .
\label{eq:rmo_final_correct}
\end{equation}

\( \mathbf{w_a} \in \mathbb{C}^{N \times 1} \) is the vectorized representation of the analog phase 
weights across all antenna elements, which were previously represented in a block-diagonal form using  \( \mathbf{W_A} \).
The matrix \( \mathbf{W_0} \in \mathbb{C}^{N \times N} \) is constructed by 
block-diagonalizing \( \mathbf{w_0} \) according to the sub-array partition. 
Equation \eqref{eq:rmo_final_correct} serves as the final optimization 
problem solved using Riemannian manifold optimization, which updates 
\( \mathbf{w_a} \) over a complex circle manifold to maximize the 
objective while preserving the constant-modulus constraint. The 
optimization on the complex circle manifold is implemented using 
MANOPT~\cite{manopt} with a steepest-gradient solver. After convergence, 
the optimized \( \mathbf{w_a} \) is reshaped into the block-diagonal 
matrix \( \mathbf{W_A} \), which can then be used in 
\eqref{eq:wd_given_WA} to compute the corresponding digital beamformer 
\( \mathbf{w_D} \).

\vspace{-2 pt}
\subsection{Practical Considerations for Realizing H-MVDR}
\label{sec:hmvdr_limitations}
In realizable HBF receivers, the observable data after analog combining yield a lower-dimensional covariance
\begin{equation}
\hat{\mathbf{R}}_D = \mathbf{W_A} \hat{\mathbf{R}}_A \mathbf{W_A}^{H}, 
\label{eq:hybrid_cov}
\end{equation}
where \( \hat{\mathbf{R}}_D \in \mathbb{C}^{N_D \times N_D} \) and  
\( \hat{\mathbf{R}}_A \in \mathbb{C}^{N_A \times N_A} \) represent the  
SCMs estimated across the digital and analog domains, respectively, over \( K \) snapshots.  
In practice, neither the \( N_A \)-dimensional ACM \( \mathbf{R}_A \) nor its SCM  \( \hat{\mathbf{R}}_A \) can be directly measured, which prevents evaluation of \eqref{eq:w0_virtual_mvdr} - \eqref{eq:rmo_final_correct} on hardware.

Prior studies implicitly assumed access to \( \mathbf{R}_A \) when formulating their optimization problems, which limits their conclusions to idealized and/or clairvoyant conditions. When the objective shifts from theoretical characterization to practical implementation, the unavailability of \( \mathbf{R}_A \) or \( \hat{\mathbf{R}}_A \) becomes a central obstacle. Such assumptions are reasonable when deriving theoretical upper bounds for S-HBF performance relative to DBFs. However, this raises an important question: how can these methods be practically realized when only the lower-dimensional digital covariance \( \hat{\mathbf{R}}_D \) is available?

The subsequent work answers that question.  The next section introduces a framework for estimating a virtual covariance \( \hat{\mathbf{R}}_{vA} \in \mathbb{C}^{N_A \times N_A} \) from \( \hat{\mathbf{R}}_D \), enabling an H-SMI beamformer that operates without full covariance access while remaining compatible with the existing optimization framework.

\section{Covariance Completion and H-SMI Formulation}

This section introduces a framework called "Rocky Road to Dublin (RR2D)", that is developed to reconstruct a estimate of the analog ACM \( \hat{\mathbf{R}}_{vA} \) required for S-HBF  when only observations from \( \hat{\mathbf{R}}_D \) are available. 

RR2D comprises two key components: (i) construction of a partially observed covariance through a hierarchical cross-sub-array switching strategy across the analog combiner (see \ref{subsec:HCSSS}), and (ii) completion of the covariance using a Toeplitz Initialization and then iteration using Dykstra’s alternating projection algorithm \cite{dykstra} with physical and structural constraints (see \ref{subsec: Dykstra}). The resulting virtual covariance estimate \( \hat{\mathbf{R}}_{vA} \) can then be substituted in lieu of \( \mathbf{R}_{A}\) into existing  H-MVDR formulation outlined in \ref{sec:hmvdr_manifold} without modifying their optimization procedures.
\vspace{-2 pt}
\subsection{Hierarchical Cross-sub-array Switching Strategy} \label{subsec:HCSSS}
In the S-HBF configuration, during each observation period, only one element per sub-array is activated while all other elements within that sub-array remain inactive.

Let each sub-array \( d \) have an index set \( \mathcal{S}_d = \{ s_{d,1}, s_{d,2}, \ldots, s_{d,N_S} \} \). At a given sample \( k \), let \( \boldsymbol{\pi}[k] \) denote the active-element configuration across all sub-arrays, defined as
\begin{equation}
    \boldsymbol{\pi}[k] = (s_{1,p_1}, s_{2,p_2}, \ldots, s_{N_D,p_{N_D}}),
    \quad p_d \in \{1,\ldots,N_S\}.
\end{equation}
For each configuration, \( K_S \) temporal snapshots are collected to form an SCM corresponding to that activation pattern.

The sequence of switching configurations is designed such that, across all snapshots, every element from one sub-array is paired \( K_S \) times with every element from all other sub-arrays. In practice, this is implemented through a hierarchical switching schedule: the last sub-array cycles through its elements first while all others remain fixed. Once all of its elements have been paired, the penultimate sub-array advances to its next element, and the process repeats until all inter-sub-array combinations have been sampled.

This switching pattern efficiently captures all inter-sub-array cross-correlations while never activating more than one element per sub-array. However, since elements within each sub-array are never active simultaneously, the intra-sub-array covariance entries remain unobserved, resulting in a structured pattern of missing entries in the analog-domain SCM.
\vspace{-3 pt}
\subsection{Matrix Completion via Structure Exploitation} \label{subsec: Dykstra}
The switching process described in the previous subsection yields an incomplete SCM, denoted \( \hat{\mathbf{R}}_{\text{inc}} \), whose missing entries correspond to intra-sub-array pairs that are never simultaneously active. The goal of this stage is to recover a feasible and physically consistent analog-domain ACM, \( \hat{\mathbf{R}}_{vA} \), that satisfies the structural and measurement constraints implied by the array geometry and the hybrid sampling process. Formally, the completion problem seeks
\begin{equation}
    \hat{\mathbf{R}}_{vA} \in 
    \mathcal{C}_{\text{PSD}} \cap 
    \mathcal{C}_{\text{T}} \cap 
    \mathcal{C}_{\text{B}},
    \label{eq:intersection}
\end{equation}
where \( \mathcal{C}_{\text{PSD}} = \{ \mathbf{R} \succeq 0 \} \) enforces positive semidefiniteness,  
\( \mathcal{C}_{\text{T}} = \{ \mathbf{R} \text{ is Hermitian Toeplitz} \} \) captures the spatial stationarity across a linear array, and  
\( \mathcal{C}_{\text{B}} = \{ \mathbf{R}_{ij} = \hat{\mathbf{R}}_{\text{inc},ij}, (i,j)\in\Omega \} \) preserves consistency with the observed SCM entries.

\paragraph*{Toeplitz Initialization}
The missing entries of \( \hat{\mathbf{R}}_{\mathrm{inc}} \) are initialized using a distance weighted Toeplitz consistent estimate, denoted \( \hat{\mathbf{R}}_{\mathrm{int}} \). Unlike standard lag averaging approaches, each missing entry is formed by combining all observed entries with the same spatial separation, with weights that decrease as a function of their distance from the target matrix index.

For each missing entry \( (i,j) \notin \Omega \), let
\[
d = |i-j|
\]
and define the set of observed entries having the same separation as
\[
\Omega_d = \left\{ (p,q) \in \Omega \;:\; |p-q| = d \right\}.
\]
Then the initialization is given by
\begin{equation}
[\hat{\mathbf{R}}_{\mathrm{int}}]_{i,j} =
\begin{cases}
[\hat{\mathbf{R}}_{\mathrm{inc}}]_{i,j}, & (i,j) \in \Omega, \\[1ex]
\displaystyle \sum_{(p,q)\in\Omega_d} w_{i,j}^{(p,q)} [\hat{\mathbf{R}}_{\mathrm{inc}}]_{p,q}, & (i,j) \notin \Omega \text{ and } \Omega_d \neq \emptyset, \\[2ex]
0.01, & (i,j) \notin \Omega \text{ and } \Omega_d = \emptyset,
\end{cases}
\end{equation}
where the weights are
\begin{equation}
w_{i,j}^{(p,q)} =
\frac{\displaystyle \frac{1}{\sqrt{(i-p)^2 + (j-q)^2} + \epsilon}}
{\displaystyle \sum_{(p',q')\in\Omega_d} \frac{1}{\sqrt{(i-p')^2 + (j-q')^2} + \epsilon}},
\end{equation}
with \( \epsilon > 0 \) a small regularization constant.

This construction promotes approximate Toeplitz structure while assigning greater influence to observed entries that are closer to the target location in the matrix. Thus, the initialization is not a simple average over entries with the same lag, but a locally weighted average over entries with the same separation. After filling the missing entries, the matrix is symmetrized as
\begin{equation}
\hat{\mathbf{R}}_{\mathrm{int}} \leftarrow \frac{1}{2}\left(\hat{\mathbf{R}}_{\mathrm{int}} + \hat{\mathbf{R}}_{\mathrm{int}}^{H}\right).
\end{equation}

For a uniformly spaced linear array under wide sense stationary signals and noise, the covariance matrix is Toeplitz \cite{Toeplitzbeex}. This initialization exploits that structure while preserving local consistency with the available entries. Since the resulting matrix is not guaranteed to be positive semidefinite, it is further refined in the subsequent projection based completion stage.

\paragraph*{Dykstra's Alternating Projection Framework}
The completion algorithm follows Dykstra’s alternating projection method, which iteratively projects the current estimate onto the convex constraint sets defined in \eqref{eq:intersection}. 
Let \( \hat{\mathbf{R}}^{(t)} \) denote the current estimate at iteration \( t \). 
Each iteration sequentially performs:
\begin{enumerate}
    \item \textbf{PSD projection:} enforce \( \hat{\mathbf{R}}^{(t)} \succeq 0 \) by eigen-decomposition and clamping negative eigenvalues to zero.
    \item \textbf{Toeplitz projection:} average entries along each diagonal to impose spatial stationarity.
    \item \textbf{Block consistency projection:} replace all entries in the observed set \( \Omega \) with their measured values from \( \hat{\mathbf{R}}_{\text{inc}} \).
\end{enumerate}
Auxiliary correction terms are updated at each step following Dykstra’s formulation to ensure convergence to a point in the intersection rather than merely oscillating among the constraint sets.
\paragraph*{Regularization and Convergence}
To mitigate numerical drift and maintain positive definiteness under limited observations, a small diagonal regularization term \( \epsilon \mathbf{I} \) is added after each PSD projection. 
The iterations continue until the relative change between successive estimates satisfies
\begin{equation}
    \frac{\|\hat{\mathbf{R}}^{(t+1)} - \hat{\mathbf{R}}^{(t)}\|_F}{\|\hat{\mathbf{R}}^{(t)}\|_F} < \delta,
\end{equation}
where \( \delta \) is a prescribed convergence tolerance. Alternatively, this could also be run until a pre-determined number of iterations are complete. 
The final estimate \( \hat{\mathbf{R}}_{vA} = \hat{\mathbf{R}}^{(t_\text{final})} \) represents a completed virtual SCM estimate that satisfies the PSD, Toeplitz, and block-consistency constraints or at the very least satisty them in a least squares sense.

This completed virtual SCM \( \hat{\mathbf{R}}_{vA} \) obtained through the proposed RR2D framework is then supplied to the H-SMI formulation introduced earlier in Section~\ref{sec:hmvdr_manifold}. 
By substituting \( \hat{\mathbf{R}}_{vA} \) for the idealized \({\mathbf{R}}_A \) in \eqref{eq:w0_virtual_mvdr} and \eqref{eq:mse_formulation}, the beamformer weights can be computed within the original Riemannian optimization framework using measurable quantities. 
This integration preserves the theoretical structure of the prior formulation while ensuring its practical realizability on hybrid architectures.

\section{Empirical Validation and Performance Analysis}
This section validates the proposed RR2D framework and compares its performance against established HBF and DBF methods. The analysis is divided into two parts: 
description of the simulation setup, and quantitative comparison of the resulting SINR performance across competing formulations.


\subsection{Nonconvexity Verification}

\begin{figure}[!t]
    \centering
    \includegraphics[
        width=\columnwidth,        keepaspectratio=false
    ]{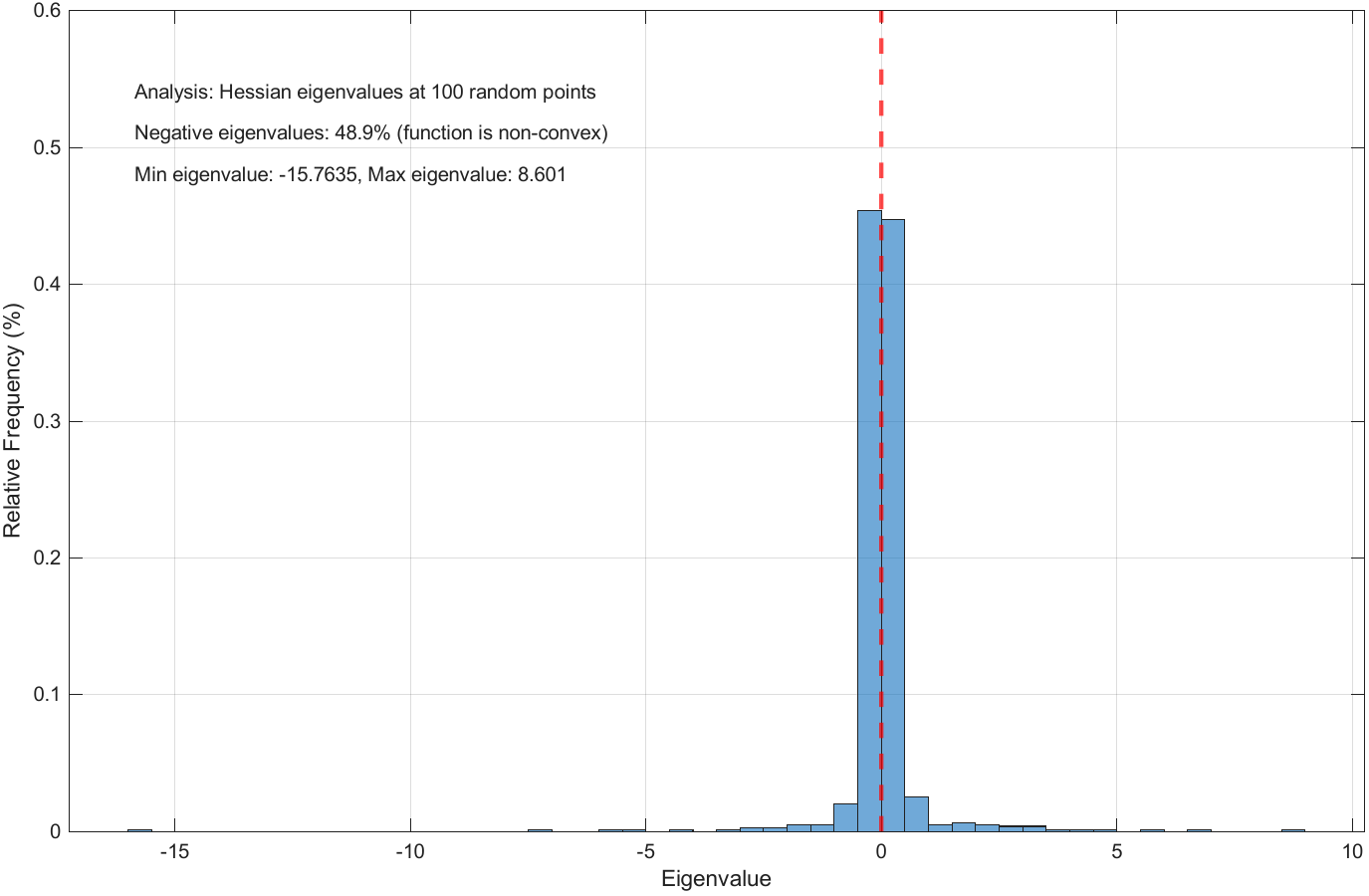}
    \caption{PDF of the Hessian eigenvalues of the objective function, evaluated at 100 randomly selected points across 1000 realizations. The distribution exhibits both positive and negative eigenvalues, indicating that the objective is neither convex nor concave over the sampled domain.} \captionsetup{font=footnotesize,skip=2pt}
    \label{fig:eig_analysis}
\end{figure}

The optimization problem defined in Section~\ref{sec:hmvdr_formulation} is nonconvex due to the bilinear coupling between the analog and digital beamforming weights and the unit-modulus constraints on the analog domain. To verify this property empirically, we compute and examine the Hessian of the objective function evaluated at randomly generated feasible points on the Riemannian manifold. In Figure \ref{fig:eig_analysis} the eigen-spectrum of the Hessian exhibits both positive and negative values, confirming the indefinite curvature of the cost landscape and the presence of multiple local minima, thus proving the problem is non-convex. 

This observation aligns with the intuition first proposed in prior work, where the unit magnitude constraint on the phase shifters between  was hypothesized to induce nonconvexity. The explicit verification here provides additional evidence of that behavior and supports the use of manifold-based optimization tools rather than convex relaxations.

\subsection{Simulation Setup and Metrics}

All simulations consider a 32-element uniform linear array partitioned into \( N_D = 2 \) digital channels/sub-arrays, each containing \( N_S = 16 \) antenna elements. Two independent RF chains are assumed, corresponding to the two sub-arrays in the hybrid configuration.

Unless otherwise specified, the received input comprises of one SOI and two interferers. Interference and SOI angles are sampled uniformly across \(\theta \in [-90,90]\), and their powers are set to yield an interference-to-noise ratio (INR) of 20~dB. The signal-to-noise ratio (SNR) is varied between \(-30\)~dB and \(30\)~dB in 2~dB increments. The choice of $N_A$, number of interferers, SOI, and INR were informed by prior work to replicate it as close as possible.

For the hybrid configuration, the switching pattern described in Section~\ref{subsec: Dykstra} is applied. Each configuration \( \boldsymbol{\pi}[k] \) is held for \( K_S = 4 \) snapshots, such  that after the sampling process is complete each array element observes a total of 64 samples. The DBF case uses the same total number of samples (\( 2N = 64 \)), ensuring an equivalent sample budget across all methods and consistency with the criterion proposed in RMB for achieving within 3~dB of the ideal MVDR performance. The DBF-SMI implementation in this work adheres to that specification.

Performance is quantified in terms of the output SINR defined as
\begin{equation}
    \text{SINR}_{\text{out}} = 
    \frac{\mathbf{w}^H \mathbf{R}_s \mathbf{w}}
         {\mathbf{w}^H \mathbf{R}_{i+n} \mathbf{w}},
\end{equation}
where \( \mathbf{R}_s \) and \( \mathbf{R}_{i+n} \) denote the signal and interference-plus-noise ACMs, respectively. Each reported point corresponds to the mean SINR across \(500\) realizations with independent noise and phase realizations.

We note that while we follow MVDR/SMI calculation outlined in \ref{sec:hmvdr_formulation}, we use the \( \mathbf{R}_{i+n} \) or \( \mathbf{\hat{R}}_{i+n} \) instead of the total ACM/SCM (\( \mathbf{R}_{s+i+n} \) or \( \mathbf{\hat{R}}_{s+i+n} \)) comprising of the SOI, interference and the noise outlined in \cite{fc-mvdr,h-mvdr}. This follows best practices \cite{dudgeon} to ensure that the signal subspace from (\( \mathbf{R}_{s+i+n} \) or \( \mathbf{\hat{R}}_{s+i+n} \)) does not leak into the precision matrix used in MVDR/SMI and null the SOI.

\subsection{Results and Discussion}
\begin{figure}[!t]
    \centering
    \includegraphics[
        width=\columnwidth,        keepaspectratio=false
    ]{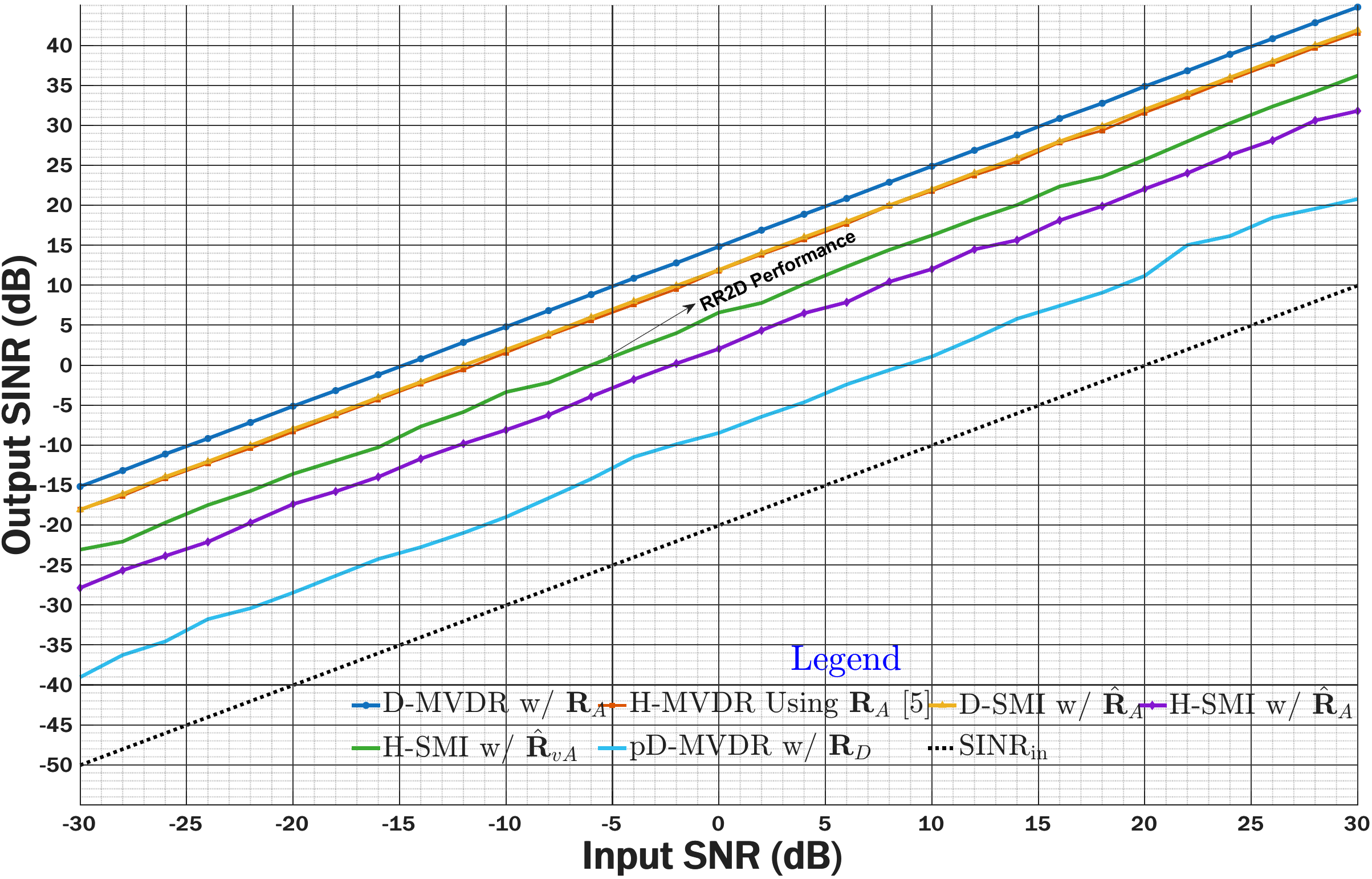}
    \caption{Mean SINR as a function of SNR for an INR of 20 dB, averaged over \(500\) randomized interference and SOI realizations for each Input SNR} \captionsetup{font=footnotesize,skip=2pt}
    \label{fig:SINR_Results}
\end{figure}

The following methods are compared in Figure \ref{fig:SINR_Results}:
 \begin{enumerate}
    \item \textbf{D-MVDR w/ \(\mathbf{R}_A\):} MVDR implemented on an \(N_A\) element DBF using \(\mathbf{R}_A\).
    
    \item \textbf{H-MVDR w/ \(\mathbf{R}_A\):} MVDR implemented on an S-HBF with \(N_D\) digital channels assuming knowledge of \(\mathbf{R}_A\) (prior work \cite{h-mvdr}).
    
    \item \textbf{D-SMI w/ \( \hat{\mathbf{R}}_A \):} SMI implemented on an \(N_A\) element DBF using \( \hat{\mathbf{R}}_A \).
    
    \item \textbf{H-SMI w/ \( \hat{\mathbf{R}}_A \):} SMI implemented on an HBF with \(N_D\) digital channels assuming knowledge of \( \hat{\mathbf{R}}_A \).
    
    \item \textbf{H-SMI w/ \(\hat{\mathbf{R}}_{vA}\):} SMI implemented on an HBF with \(N_D\) digital channels using the RR2D framework to reconstruct \(\hat{\mathbf{R}}_{vA}\) from \(\hat{\mathbf{R}}_{D}\) prior to steps (5) to (9).
    
    \item \textbf{pD MVDR w/ \({\mathbf{R}}_{D}\):} MVDR implemented on a P-DBF with \(N_D\) elements/digital channels using \({\mathbf{R}}_{D}\) (baseline in \cite{h-mvdr})..
\end{enumerate}

Figure~\ref{fig:SINR_Results} shows the output SINR as a function of the input SNR for all methods. The input SINR (dotted black line), serves as a reference for comparison. As expected, the fully digital MVDR with access to the true ACM \( \mathbf{R}_A \) (plotted in dark blue) performs best and is henceforth called the \emph{Oracle}. The Oracle achieves an average SINR improvement of approximately 35~dB relative to the input SINR. 

Results from the H-MVDR with knowledge of \( \mathbf{R}_A \) (plotted in gold) and the H-MVDR with knowledge of \( \hat{\mathbf{R}}_A \) (plotted in red) closely follow each other, achieving around 31–32~dB of improvement over the input SINR, or roughly 3~dB below the Oracle. Notably, the performance of H-MVDR trails the Oracle far more closely than reported in prior work~\cite{h-mvdr} where it was first introduced. We attribute this to the use of a trust-region solver rather than simple steepest gradient solver, as well as to averaging over \(500\) Monte Carlo realizations, which provides a more stable statistical estimate. The H-SMI computed on the Riemannian manifold suffers a noticeable drop (between 12.4-13.03 dB) in SINR relative to the Oracle. The H-SMI (purple) approach that leverages the virtual ACM reconstructed through the RR2D framework (green) performs between the H-SMI with knowledge of \( \mathbf{R}_A \) and the H-SMI that relies directly on the SCM \( \hat{\mathbf{R}}_A \). When compared to the Oracle, the H-SMI approach using the virtual covariance matrix notices a SINR drop of ~8.3-9 dB.  Additional experiments (not shown) revealed that removing the Toeplitz constraint from the Dykstra's iteration reduced the achieved SINR below that of the H-SMI using \( \hat{\mathbf{R}}_A \), confirming that the Toeplitz projection is critical for statistical regularization and robust weight computation. The P-DBF MVDR (cyan) using \( \mathbf{R}_D \) performs better than the input SINR (by~12-13 dB) but struggles to simultaneously suppress both interferers. 

In summary, the proposed RR2D-enabled H-SMI, (i) enables a practical means of performing H-SMI using only $N_D$ digital channels, (ii) maintains a SINR gap of 8.3-9.2 dB in most SNR regimes relative to the Oracle. These results confirm that the proposed framework extends prior hybrid MVDR formulations to practical finite-snapshot scenarios and improves robustness to partial observations.



\section{Limitations, Future Work, and Discussion}
The present work presents a viable path to implementing SMI on S-HBFs, but, it also carries several important limitations. In practical implementations, array geometries may be nonuniform or irregular, and the assumption of a Toeplitz covariance structure may not hold. Strategies tailored to specific array geometries must be developed and exploited accordingly. This work also does not account for the bandwidth of either the signal of interest or the interferers, a simplification that limits applicability to real-world wideband systems. Future work can extend the proposed framework by incorporating bandwidth effects during covariance generation and reconstruction.

Practical arrays are additionally subject to phase quantization, ADC quantization noise, phase noise, and mutual coupling, none of which are modeled here. Incorporating these hardware impairments directly into the optimization framework would yield a more accurate and robust design. Moreover, while the current study assumes Gaussian noise and uses the SCM as an estimator, M-estimators and other robust covariance estimation techniques should be investigated for non-Gaussian environments.The sampling strategy used in this work enforces an identical number of samples across all array elements for both the DBF and RR2D frameworks. The additional time required for hierarchical cross-subarray switching may limit applicability in time-sensitive sampling scenarios. Future studies should investigate the trade-offs introduced by sampling over fixed time intervals and explore whether adopting a sparse-ruler-based sampling scheme could reduce the number of required element activations while maintaining comparable performance.

The proposed covariance completion strategy can also be extended to direction-of-arrival (DOA) estimation and related hybrid array processing tasks. Such extensions would require adapting the RR2D framework to preserve structural properties relevant to spatial spectrum estimation. 

Finally, the convergence behavior and computational scaling of Dykstra’s algorithm under large arrays or limited observations remain to be characterized. The sensitivity of RR2D performance to the choice of subarray configuration and switching schedule also warrants further study.

\section{Conclusion}
This work demonstrated a practical realization of SMI for sub array based hybrid beamforming architectures by combining Toeplitz constrained covariance completion with a Riemannian optimization framework. The proposed RR2D approach enables reconstruction of a full dimensional covariance matrix from low dimensional hybrid observations and supports both MVDR and SMI processing without requiring modifications to the analog front end. Simulation results showed that the recovered covariance matrices produce beamformers whose performance approaches the performance of H-MVDR approaches that have access to the ACM, under a wide range of SNR and interference conditions

\section*{Acknowledgment}
This research was partially supported by the U.S. Department of Commerce’s National Telecommunications and Information Administration (NTIA) under the Public Wireless Supply Chain Innovation Fund Grant Program (Award 24-60-IF2415: ASPEN - Advanced Signal Processing Enhancement for Next-Generation Open Radio Units), administered by the National Institute of Standards and Technology. This research was partially supported by Science of Tracking, Control, and Optimization of Information Latency for Dynamic Military IoT Systems
Office of Naval Research (N00014-19-1-2621). This research was partially supported by Learning to Prevail: Communication in Contested and Adversarial Environments AFRL prime FA8750-20-2-0504. This work was developed on a NVIDIA DGX Platform provided by Cambridge Computer. We thank Analog Devices, MathWorks and Rohde and Schwarz for hardware and software support provided during the duration of this work. 

\bibliographystyle{IEEEtran}
\bibliography{references.bib} 

\end{document}